\documentclass[12pt]{article}
\usepackage[latin1]{inputenc}
\usepackage[T1]{fontenc}
\usepackage[english]{babel}
\usepackage{amssymb}
\usepackage{amscd}
\usepackage{fancyhdr}
\usepackage{color}
\usepackage{epsfig}
\usepackage{graphicx}
\usepackage{bm}
\usepackage{subfigure}
\definecolor{red}{rgb}{1., 0., 0.}
\definecolor{blue}{rgb}{0., 0., 1.}
\definecolor{green}{rgb}{0.1, 0.7, 0.}
\definecolor{purp}{rgb}{1,0,1}
\newcommand{\bc}{\begin{cases}\begin{aligned}}
\newcommand{\ec}{\end{aligned}\end{cases}}

\newcommand{\eq}{\begin{equation}}
\newcommand{\fine}{\end{equation}}

\def\drawbox#1#2{\hrule height#2pt
        \hbox{\vrule width#2pt height#1pt \kern#1pt
              \vrule width#2pt}
              \hrule height#2pt}

\def\Asym#1#2{\vcenter{\vbox{\drawbox{#1}{#2}
              \kern-#2pt       
              \drawbox{#1}{#2}}}}

\newcommand {\beq} {\begin{equation}}
\newcommand {\eeq} {\end{equation}}
 \newcommand{\be}{\begin{eqnarray}}
\newcommand{\ee}{\end{eqnarray}}

\begin{document}

\begin{titlepage}

\begin{flushright}
\small ZMP-HH/05-22 \
\end{flushright}

\vspace {1cm}
\centerline{\Large \bf Fractional branes and} 
\vspace {.4cm}
\centerline{\Large \bf the gravity dual of partial supersymmetry breaking}

\vskip 1cm \centerline{\large Paolo Merlatti} \vskip 0.1cm

\vskip .5cm

\vskip 0.5cm \centerline{II. Institut f\"ur Theoretische Physik Universit\"at Hamburg,} \centerline{Luruper Chaussee 149 D-22761 Hamburg, Germany}\centerline{ merlatti@mail.desy.de}

\vskip 1cm

\begin{abstract}
In type IIB string theory, we consider fractional $D3$-branes in the orbifold background dual to four-dimensional ${\cal N}=2$ supersymmetric Yang-Mills theory. We find the gravitational dual description of the generation of a non-trivial field theory potential on the Coulomb branch. When the orbifold singularity is softened to the smooth Eguchi-Hanson space, the resulting potential induces spontaneous partial supersymmetry breaking. We study the ${\cal N}=1$ theory arising in those vacua and we see how the resolved conifold geometry emerges in this way. It will be natural to identify the size of the blown-up two-cycle of that geometry with the inverse mass of the adjoint chiral scalar. We finally discuss the issue of geometric transition in this context.   
\end{abstract}

\end{titlepage}
\tableofcontents

\section{Introduction}

In recent years supersymmetric Yang-Mills theories have been much investigated and a number of non-trivial results about their non-perturbative dynamics have been reached. The motivation of this intensive investigation is string theory. Indeed it has become clear that the embedding of gauge theories in string theory (via $D$-branes) is a powerful tool to explore various interesting features of their dynamics. Within this framework, several approaches have been developed, all going under the rather generic name of gauge/string correspondence. The most celebrated example is the ${\cal N}=4$ AdS-CFT duality \cite{Maldacena:1997re} (see \cite{Aharony:1999ti} for an excellent review). Many interesting results have been also found for less (super)-symmetric theories (see \cite{Aharony:2002up} for some extensive reviews). In this paper we make contact with a nice and fruitful way of looking at this duality. Such perspective has been introduced in \cite{Gopakumar:1998ki} for the topological string. There, it has been shown that in topological string theory the deformed conifold background, with $N$ 3-brane wrapped on $\mathbf{S}^3$ (equivalent to the $\mathbf{S}^3$ Chern-Simon theory \cite{Witten:1992fb}), is dual to the same topological string but without string modes and on a different background. The new background is the resolved conifold, whose parameters depend now on $N$.
Such gauge/geometry correspondence has been successfully embedded in superstring theory in \cite{Vafa:2000wi}. The central idea is the concept of geometric transition. The field theory has to be engineered via $D$-branes wrapped over certain cycles of a non-trivial Calabi-Yau geometry. The low energy dual arises from a geometric transition of the Calabi-Yau itself, where the branes have disappeared and have been replaced by fluxes. The simplest example \cite{Vafa:2000wi,Cachazo:2001jy} is pure ${\cal N}=1$ super Yang-Mills theory. Before the transition the field theory is engineered on $N$ $D5$ branes wrapped on the blown-up $\mathbf{ S}^2$ of a resolved conifold. After the transition we are instead left with $N$ units of R-R flux through the $\mathbf{S}^3$ of a deformed conifold \cite{Klebanov:2000hb}.

In this paper we discuss the gauge/string correspondence starting from ${\cal N}=2$ supersymmetric theories and we find how partial supersymmetry breaking can be described on the supergravity side. Within this purely ten dimensional superstring context, we see then how the resolved conifold background emerges naturally as the proper one to engineer ${\cal N}=1$ super-Yang-Mills theory. 

In a first part we discuss the string dual of ${\cal N}=2$ supersymmetric $SU(N)$ Yang-Mills theories with non-trivial potential for the adjoint scalar. We move from the general observation that the same supergravity solution can be given in terms of forms of different degree, that are related to each other by electric-magnetic duality. If we consider for example type II supergravity, the same $Dp$-brane solution can be written with an electric ansatz (the relevant R-R field in the solution is the potential $C_{p+1}$) or with a magnetic one (the relevant R-R potential is $C_{7-p}$). The two solutions should be equivalent. Indeed they are, from a bulk perspective. It is perhaps surprising that, as we show in this paper for the specific ${\cal N}=2$ case, they correspond to different gauge dynamics of the dual Yang-Mills theory. To see this, we add probe-branes in the background geometry, and the boundary action describing the Yang-Mills theory living on the world-volume of such branes will give different results according to which kind of ansatz (electric or magnetic) one makes for the bulk solution.

To be more specific, we consider the orbifold background $\mathbb{C}_2/\mathbf{Z}_2$ with just one kind of fractional branes and no bulk ones. The complete explicit bulk solution for this ${\cal N}=2$ supersymmetric background has been found in \cite{Bertolini:2000dk}. Such solution can be described equivalently in terms of the R-R twisted four-form potential $A_4$ or of its six-dimensional magnetic dual, the twisted scalar $c$. In this paper we study the dual Yang-Mills theory and we find that, according to which bulk field is used to describe the solution, its low energy properties are different. Making a probe analysis, we find that if the solution is given in terms of $A_4$ (electric ansatz), there is no potential for the $U(1)$ gauge theory living on the world-volume of the probe. Contrary, if the solution is given in terms of $c$ (magnetic ansatz), a non-trivial potential is generated. Thus, the magnetic ansatz corresponds to switching on an internal flux that generates a gauge theory potential. We will see 
such potential corresponds to the generation of electric-magnetic Fayet-Iliopoulos terms of the kind discussed in \cite{Antoniadis:1995vb}. Even if both the results are consistent with ${\cal N}=2$ quantum field theory, we find it surprising that the Yang-Mills dual physics depends crucially on the kind of ansatz one makes to describe the bulk geometry. In this paper this observation is just empirical, but this phenomenon could be more general and it deserves further investigation. We are indeed currently studying the origin of such discrepancy via a detailed stringy analysis.

In the second, more speculative, part of the paper we see directly in type II superstring theory how the resolved conifold geometry emerges naturally as the proper geometry to engineer ${\cal N}=1$ supersymmetric Yang-Mills theories. This part is a development of the first one. Starting in fact from the orbifold background, particularly  
from the analysis we make for the magnetic case, we consider its generalization to the case where the singular orbifold geometry is softened to the smooth Eguchi-Hanson space. We find vacua where partial supersymmetry breaking occurs and we study the description of ${\cal N}=1$ super-Yang-Mills theory in such vacua. This study leads us to argue that the natural background where ${\cal N}=1$ Yang-Mills theories should be engineered is the resolved conifold. Besides we gain information on the way such engineering works. In particular it turns out to be quite natural to identify the mass of the chiral adjoint superfield with the inverse of the size of the blown-up two-cycle. Shrinking this two-cycle to zero size corresponds thus to decouple the adjoint scalar and to get pure ${\cal N}=1$ super-Yang-Mills in the singular conifold background. We know eventually the rest of the story: to cure the singularity of this geometry, the ${\mathbf S}^3$ has to be blown up \cite{Vafa:2000wi,Klebanov:2000hb}. This corresponds to generate the infrared dynamically generated scale of the dual gauge theory (in this case coming from gaugino condensation).
  
The picture we get in this way has an immediate translation in the type IIA T-dual scenario of rotated NS-branes \cite{Barbon:1997zu,Uranga:1998vf,Dasgupta:1999wx}. We will comment on this relation in section 4.

Another way to make contact with the ${\cal N}=1$ conifold theory starting from a more supersymmetric theory in the ultraviolet is to start from a quiver ${\cal N}=2$ supersymmetric field theory \cite{Klebanov:1998hh}. In that case one starts from the orbifold background $\mathbf{S}^5/\mathbb{Z}_2$, that is dual to the ${\cal N}=2$ super-conformal Yang-Mills $U(N)\times U(N)$ theory with hypermultiplets transforming in $(N,\bar{N})\oplus (\bar{N},N)$. From an ${\cal N}=1$ perspective, the hypermultiplets correspond to chiral multiplets and other chiral multiplets in the adjoint representations of the two $U(N)$'s are present. There is also a non-trivial superpotential for the multiplets in the bi-fundamental representations of the gauge group. One can add now a relevant term to this superpotential. To integrate out the adjoint multiplets in presence of this relevant perturbation corresponds to blow up the orbifold singularity of $\mathbf{S}^5/\mathbb{Z}_2$ \cite{Klebanov:1998hh}. Remarkably, in \cite{Klebanov:1998hh} it has also been shown that such blown-up space is topologically equivalent to the coset space $T^{1,1}$, the base of the conifold. This way of flowing to the conifold has some points in common with our construction. The main difference is that we start from a simpler gauge theory (we have no matter and just one gauge group from the very beginning) and we follow all the steps in terms of explicit supergravity solutions.

\section{Brief review of ${\cal N}=2$ supersymmetric\\ Lagrangians}

In this section we briefly review the field theory analysis of ${\cal N}=2$ supersymmetric gauge theories (with gauge group $U(1)$), emphasizing the role of electric and magnetic Fayet-Iliopoulos (FI) terms \cite{Antoniadis:1995vb}.
The general form of a ${\cal N}=2$ Lagrangian is determined by the analytic prepotential ${\cal F}({\cal A})$ (${\cal A}$ being the ${\cal N}=2$ vector superfield):
 \begin{equation}\label{L2}
L_0~=~\frac{i}{4}\int d^2\theta_1 d^2\theta_2 {\cal F}({\cal A})~+~c.c.\end{equation}
In ${\cal N}=1$ language, the abelian vector multiplet ${\cal A}$ contains the ${\cal N}=1$ gauge multiplet $A$ and a neutral chiral superfield $\Phi$, whose complex scalar component we denote by $a$. In ${\cal N}=1$ superspace the Lagrangian (\ref{L2}) reads:
\begin{equation}
\label{generic2}
L_0~=~\frac{1}{4}\int d^2\theta f(\Phi)~ {\cal W}^2 ~+~ c.c. ~+~\int d^2\theta d^2\bar{\theta}~ K(\Phi,\bar{\Phi})
\end{equation}
where ${\cal W}$ is the standard gauge field strength superfield and
\begin{equation}
K(a,\bar{a})~=~\frac{i}{2}(a~\bar{{\cal F}}_{\bar{a}}-\bar{a}~{\cal F}_a)\hspace{1.6cm}f(a)~=~-i{\cal F}_{aa}
\end{equation}where the $a$ ($\bar{a}$) sub-script denotes derivative with respect to $a$ ($\bar{a}$).
It is well known that this Lagrangian can be supplemented by a FI term that preserves ${\cal N}=2$ supersymmetry \cite{Fayet:1975yi}:
\begin{equation}
\label{FID}
L_D~=~\sqrt{2}\xi D,
\end{equation}
with $\xi$ a real constant. More generally, it has been shown in \cite{Antoniadis:1995vb} that the following superpotential
\begin{equation}\label{supot1}
W~=~e a~+~m{\cal F}_a
\end{equation}
is also compatible with ${\cal N}=2$ supersymmetry. The resulting scalar potential is:
\begin{equation}
V_{{\cal N}=1}~=~\frac{|e~+~m\tau|^2~+~\xi^2}{\tau_2},
\label{pot1}\end{equation}
where $\tau\equiv\tau_1+i\tau_2\equiv {\cal F}_{aa}(a)$.

For $m=0$ the superpotential (\ref{supot1}) is equivalent to a FI term (\ref{FID}) with $\xi=e$ \cite{Fayet:1975yi}. It is now quite natural to interpret the term $m{\cal F}_a$ in (\ref{supot1}) as a magnetic FI term (it is related to the electric one by $S$ duality). In \cite{Antoniadis:1995vb} it has been shown how such interpretation becomes even more clear if a manifestly ${\cal N}=2$ supersymmetric formalism is used. In that case the field theory analysis shows also that a constant shift to the potential (\ref{pot1}) is still compatible with ${\cal N}=2$ supersymmetry. The resulting potential is: 
\begin{equation}V~=~V_{{\cal N}=1}~+~2~m~p~=~\frac{|e~+~m\tau|^2~+~\xi^2}{\tau_2}~+~2~m~p,\label{pot2}\end{equation}where $p$ is an arbitrary real constant.

\subsection{Stable vacua}\label{vacua}

Again following \cite{Antoniadis:1995vb}, we now turn to the minimization of the scalar potential (\ref{pot2}). We find a vacuum at \begin{equation} \label{min}\langle\tau_1\rangle~=~-\frac{e}{m}~.\ \ \ \ \ \langle\tau_2\rangle~=~\left|\frac{\xi}{m}\right|,\end{equation}
From this relations a vacuum expectation value for the scalar field ($\langle a \rangle$) can be found. Now there are different cases to consider. Two of them are specially interesting for us: \begin{enumerate} \item $\xi m\neq 0$: in this case at the point (\ref{min}) we end up with an ${\cal N}=1$ vacuum. Therefore partial supersymmetry breaking does occur and we get here a massless $U(1)$ vector multiplet plus a massive chiral multiplet (with mass $\frac{m^2}{2\xi}\langle\tau_a\rangle$).
\item $m\neq 0,\ \xi=0$: in this case the metric $K_{a\bar{a}}\equiv\langle\tau_2\rangle$ is singular. The singularity is due to the appearance of a new massless dyonic state with quantum numbers $(m_0,e_0)$ at $\langle a\rangle$. It can further be shown that \begin{equation} (m,~e)~=~c~(m_0,~e_0),\end{equation} for some constant $c$. This dyon does condense, breaking the $U(1)$ gauge symmetry while ${\cal N}=2$ supersymmetry is unbroken. 
\end{enumerate}

\section{Orbifold geometry}

In the framework of the gauge-string correspondence it is possible to give dual supergravity descriptions of ${\cal N}=2$ supersymmetric Yang-Mills theories \cite{Bertolini:2000dk,Kachru:1998ys,Polchinski:2000mx,Gauntlett:2001ps}. The approach we follow here is to study a stack of $N$ fractional D3-branes in the IIB orbifold background (${\mathbb R}^{1,5}~\times~{\mathbb C}_2/{\mathbf Z}_2$) \cite{Bertolini:2000dk,Polchinski:2000mx}. On their world volume a ${\cal N}=2$ supersymmetric $SU(N)$ Yang-Mills theory lives. Non-trivial information on such theory can be extracted from a low energy ({\it i.e.} supergravity) analysis of this background \cite{Bertolini:2000dk,Polchinski:2000mx}. 

We start reviewing the solution found in \cite{Bertolini:2000dk}\footnote{We refer to that paper also for the notation} (with slightly different boundary conditions) and then move to its field theoretic interpretation. We will find some results that have been overlooked in the past and we will show how they are compatible with the generation of an ${\cal N}=2$ potential of the kind we discussed in the previous section.

The solution we are discussing can be written in terms of the metric, a R-R four form potential ($C_4$), a NS-NS twisted scalar ($\tilde{b}$) and a R-R twisted one ($c$):
\begin{eqnarray}\nonumber
ds^2&=&H(r,\rho)^{-1/2}\eta_{\alpha\beta}dx^{\alpha}dx^{\beta}~+~H(r,\rho)^{1/2}\delta_{ij}dx^idx^j,\\\label{orbi}  C_4&=&\left(H(r,\rho)^{-1}-1\right)~dx^0\wedge\ldots\wedge dx^3,\\c&=&~NK~\theta\hspace{1.2cm},\hspace{1.2cm}\tilde{b}~=~NK\log\frac{\rho}{\epsilon}~,\nonumber
\end{eqnarray} where $\alpha,\beta=0,\ldots,3;\ i,j=4,\ldots,9;\ r=\sqrt{\delta_{ij}x^ix^j};\ K=4\pi g_s\alpha';\ \epsilon$ is a regulator; $\rho$ and $\theta$ are polar coordinates in the plane $x^4,\ x^5$. The self-duality constraint on the R-R five-form field strength has to be imposed by hand. 

The precise functional form of the function $H(r,\rho)$ has also been determined \cite{Bertolini:2000dk} but it is not relevant for our purposes. It is enough for us to note that the spacetime (\ref{orbi}) has a naked singularity. This is a quite general feature of those gravitational backgrounds that are dual to non-conformal gauge theories. Luckily, in many cases the singularities are not actually present but they are removed by various mechanisms. In theories ${\cal N}=2$ supersymmetric the relevant mechanism is the enhan\c{c}on \cite{Johnson:1999qt}. It removes a family of time-like singularities by forming a shell of (massless) branes on which the exterior geometry terminates. As a result, the interior singularities are ``excised'' and the spacetime becomes acceptable, even if the geometry inside the shell has still to be determined. For the case at hand, the appearance of such shell, its field-theoretic interpretation and its consistency at the supergravity level have been discussed in \cite{Bertolini:2000dk,Polchinski:2000mx,Merlatti:2001gd}. It turns out that the enhan\c{c}on shell is simply a ring in the $(\rho,\theta)$-plane. Its radius is \begin{equation}\label{re}\rho_e=\epsilon~ e^{-\pi/(2Ng_s)}.\end{equation} The geometry (\ref{orbi}) is thus valid outside the enhan\c{c}on shell, {\it i. e.} for $\rho\geq\rho_e$. On the gauge theory side the enhan\c{c}on locus corresponds to the locus where the Yang-Mills coupling constant diverges \cite{Bertolini:2000dk,Polchinski:2000mx}.

\subsection{Electric case}

The solution (\ref{orbi}) can also be written following an electric ansatz, {\it i.e.} in terms of the R-R twisted field $A_4$, the field to which a fractional $D3$-brane couples electrically. $A_4$ replaces its six dimensional magnetic dual $c$. Following \cite{Bertolini:2000dk}, it is easy to determine it:
\begin{equation}\label{RRTbc}
A_4~=~NK\log\frac{\rho}{\epsilon}dx^0\wedge\ldots\wedge dx^3~+~2\pi^2\alpha'~=~NK\log\frac{\rho}{\rho_e}dx^0\wedge\ldots\wedge dx^3,
\end{equation}
where, with respect to \cite{Bertolini:2000dk}, we have changed the boundary conditions in such a way that $A_4$ is positive for every value of $\rho>\rho_e$.
 
The world-volume action of a fractional D3-brane in terms of such field reads:
\begin{eqnarray}\label{boundary}
S_{bdy}&=&-\frac{T_3}{\sqrt{2}k_{orb}}\int d^4x\sqrt{-det~G_{\alpha\beta}}\left(1+\frac{1}{2\pi^2\alpha'}\tilde{b}\right)\\ \nonumber &&+\frac{T_3}{\sqrt{2}k_{orb}}\int C_4\left(1+\frac{1}{2\pi^2\alpha'}\tilde{b}\right)+\frac{T_3}{\sqrt{2}k_{orb}}\frac{1}{2\pi^2\alpha'}\int A_4,
\end{eqnarray}
where $ T_3~=~\sqrt{\pi}$ and $k_{orb}~=~(2\pi)^{7/2}g_s\alpha'^2$.
Substituting the classical solution in the boundary action we can make the so-called probe computation (for a review of this method see, for example, \cite{Johnson:2000ch}) and get the static result 
\begin{eqnarray}\label{probeaction1}
 S_{probe}&=& -\frac{T_3V_4}{\sqrt{2}k_{orb}}\Bigg\{H^{-1}\left(1+\frac{K N}{2\pi^2\alpha'}\log\frac{\rho}{\epsilon}\right)-\left(H^{-1}-1 \right)\cdot\\&&\hspace{-0.6cm}\cdot\left(1+\frac{K N}{2\pi^2\alpha'}\log\frac{\rho}{\epsilon}\right)-\frac{1}{2\pi^2\alpha'}\left(KN\log\frac{\rho}{\epsilon}+2\pi^2\alpha'\right)\Bigg\}~=~0~\nonumber,
\end{eqnarray}
where $V_4$ is the four dimensional volume filled by the brane. The fact that all position dependent terms exactly cancel indicates that there is no potential felt by the probe-brane. The no-force condition is thus respected to all orders. This was expected because of the BPS properties of the system. 

When we identify the coordinates $x^a=2\pi\alpha'\Phi^a$ with the chiral scalar field of the Yang-Mills theory living on the world-volume of the brane, the vanishing result of (\ref{probeaction1}) corresponds to have no potential in the gauge model. This is consistent with the field theory analysis of the previous section, being $V=0$ a special case of the most general ${\cal N}=2$ potential in (\ref{pot2}), ({\it i.e.} with $e=m=\xi=0$). A constant (different from zero) result, as it has been obtained in \cite{Bertolini:2000dk}, would be an equally good check of the no-force condition, but according to (\ref{pot2}) it would not be consistent with ${\cal N}=2$ supersymmetry. The fact we get precisely zero is thus a check of the boundary condition chosen in (\ref{RRTbc}). 

Following \cite{Bertolini:2000dk}, we can also include the world-volume fields fluctuations and study the motion of the probe-brane in the geometry (\ref{orbi}). This is done by inserting the classical solution in the world-volume action (\ref{boundary}) and expanding it in powers of the velocity of the probe $D3$-brane in the two transverse directions $x^4$ and $x^5$. We find that the terms quadratic in the velocity of the probe survive and allow to define a non-trivial two-dimensional metric on the moduli space. It is easy to see \cite{Bertolini:2000dk} that this non-trivial metric corresponds holographically to the one-loop running of the Yang-Mills coupling constant, that in terms of the gravitational degrees of freedom reads:
\begin{equation}
\frac{1}{g_{Y}^2}~=~\frac{T_3}{\sqrt{2}k_{orb}}\frac{(2\pi\alpha')^2}{2}\left(1+~\frac{1}{2\pi^2\alpha'}\tilde{b}\right)\label{running}.
\end{equation}

\subsection{Magnetic case}

We now repeat the same probe analysis directly in terms of the fields appearing in (\ref{orbi}), {\it i.e.} without dualizing the R-R twisted scalar $c$, the field to which a fractional $D3$-brane couples magnetically. We should thus consider (instead of (\ref{boundary})) the following boundary action
\begin{eqnarray}
\label{boundary1}
S_{bdy}&=&-\frac{T_3}{\sqrt{2}k_{orb}}\int d^4x\sqrt{-det~G_{\alpha\beta}+2\pi\alpha'F_{\alpha\beta}}\left(1+\frac{1}{2\pi^2\alpha'}\tilde{b}\right)\\ \nonumber &&+\frac{T_3}{\sqrt{2}k_{orb}}\int C_4\left(1+\frac{1}{2\pi^2\alpha'}\tilde{b}\right)+\frac{T_3}{\sqrt{2}k_{orb}}\frac{\alpha'}{2}\int c~F\wedge F.
\end{eqnarray}
where we have also included the world-volume vector fields fluctuations. We make now the probe computation with such boundary action and get:

\begin{eqnarray}
 S_{probe}&=&-\frac{T_3}{\sqrt{2}k_{orb}}\Bigg\{\int d^4x\frac{(2\pi\alpha')^2}{2}\left(\frac{1}{2}\delta_{ab}\partial^{\alpha}\Phi^a\partial_{\alpha}\Phi^b+\frac{1}{4}F_a^{\alpha\beta}F^a_{\alpha\beta}\right)\cdot\\ &&\hspace{-1.3cm}\cdot\left(1+~\frac{NK}{2\pi^2\alpha'}\log\frac{\rho}{\epsilon}\right) + V_4\left(1+\frac{NK}{2\pi^2\alpha'}\log\frac{\rho}{\epsilon}\right)-\frac{\alpha'}{2}KN\theta\int F\wedge F\Bigg\}.\nonumber\label{probeaction2}
\end{eqnarray}
One immediate consequence of doing the probe computation in this way is that we respect holomorphicity getting also the expected term reproducing the one-loop chiral anomaly \cite{Bertolini:2002xu}. 
Besides we see also that a non trivial potential is generated, namely:
\begin{equation}
V(\Phi)~=~\frac{T_3}{\sqrt{2}k_{orb}}\left(1+\frac{NK}{2\pi^2\alpha'}\log\frac{\rho}{\epsilon}\right)~=~\frac{1}{(2\pi)^3\alpha'^2}~ \tau_{Y2}\label{potm}
\end{equation}
where, consistently with the electric result (\ref{running}), we have defined $\tau_Y\equiv\tau_{Y1}+i\tau_{Y2}\equiv \frac{4\pi i}{g^2_Y}+\frac{\theta_Y}{2\pi}$ as the holographic complexified Yang-Mills coupling \cite{Bertolini:2002xu}:
\begin{equation}\label{varie}\tau_{Y}~=~\frac{1}{(2\pi\sqrt{\alpha'})^2g_s}\left( c+i\left(2\pi^2\alpha'+\tilde{b}\right)\right).\end{equation}

This computation shows that, when the solution is written in term of a magnetic ansatz, the no-force property of the BPS system is lost. This could seem quite surprising as the configuration is still ${\cal N}=2$ supersymmetric and usually the no-force condition is respected with such amount of supersymmetry (on the field theory side such condition reflects the flatness of the Coulomb branch). Nevertheless, at least from a purely field theoretical point of view, we have seen in the previous section that such flatness can be violated by electric-magnetic Fayet-Iliopoulos terms and as a result the potential (\ref{pot1}) is still compatible with ${\cal N}=2$ supersymmetry. Luckily, it turns out the potential (\ref{potm}) is just a specific case of (\ref{pot1}), namely with \begin{equation}e~=~-\frac{m\theta_Y}{2\pi}\ ,\ \ \ \xi=0\ \ \ \mbox{and}\ \ \ m^2~=~\frac{1}{(2\pi)^3\alpha'^2}\label{values}.\end{equation} We see in this way that the violation of the no-force condition does not contradict supersymmetry, but it corresponds to the generation of a purely magnetic FI term. We call it purely magnetic because a dyon with quantum numbers $(e,m)$ has an effective electric charge equal to $e+\frac{m\theta}{2\pi}$ \cite{Witten:1979ey} and with the values in (\ref{values}) it vanishes identically. Such values ensure also that the probe-brane is in a proper vacuum in the $\hat{\theta}$ direction, but for generic values of $\rho$ it is not and it moves toward $\rho=\rho_e$. This situation holographically corresponds to the second type of vacua discussed in section (\ref{vacua}), where we have learned it preserves all the supersymmetries but it signals a singularity. This continues to be in perfect agreement with the gravity picture we just got: the probe-brane is pushed to the enhan\c{c}on and there stays. From the gauge theory it has been moreover possible to argue at that point of the moduli space extra massless states appear (i.e. monopoles) and to give a physical interpretation of the singularity. On the gravity side it corresponds to the well known fact that at the enhan\c{c}on extra massless objects appear \cite{Johnson:1999qt} and it enforces the interpretation of the enhan\c{c}on locus as the field theoretic curve of marginal stability (we mean the curve in moduli space where extra massless objects, typically dyons, appear \cite{Ferrari:1996sv}).

\section{Away from the orbifold limit} 

In this section we soften the orbifold singularity and consider the resulting (smooth) Eguchi-Hanson space \cite{Eguchi:1978xp}. We generalize the probe analysis that we have done in the orbifold for the magnetic case. We find partial supersymmetry breaking. We are then led to consider ${\cal N}=1$ supersymmetric Yang-Mills theory and the resolved conifold geometry.

This could look at first sight surprising. It is indeed well known that, in presence of fractional branes, it is not possible to soften the orbifold geometry to a smooth ALE (asymptotically locally Euclidean) space in a supersymmetric way. A nice way to understand this statement is to go to the T-dual type IIA set-up \cite{Dasgupta:1999wx}, where the configuration corresponding to fractional branes on the orbifold is that of two parallel NS5-branes with a stack of $D4$-branes stretched between them. The blowing up of the orbifold corresponds to moving the NS5-branes in the directions transverse to the $D4$-branes and this cannot be done without some cost in energy.

In our case there is a caveat to this no-go theorem. In the magnetic case we have discussed in section (3.2), the fractional brane is not at the orbifold point (background $B_{NS}$ field flux through the cycle equal to 1/2 in string units), but it sits at the singular point (with zero flux) where the no-go theorem \cite{Dasgupta:1999wx} does not apply.

\subsection{Eguchi-Hanson space}

The dual field theory analysis of the previous section has been entirely performed in an orbifold background and we have found, in the magnetic case, purely magnetic FI terms. It is well known \cite{Douglas:1996sw} that a way to introduce standard ({\it i.e.} electric) FI terms is via the blow up of the orbifold singularity. In such case the resulting space is no longer an orbifold but it is a (smooth) Eguchi-Hanson ($EH$) space \cite{Eguchi:1978xp}. This is an ALE space. It is characterized by three moduli that correspond to three different ways of blowing-up the relevant two cycle when going away from the orbifold limit (via the triplet of massless NS-NS twisted scalars). Analogously to the field theory description of ${\cal N}=2$ FI terms, it is possible to use the global $SU(2)_R$ symmetry and to reduce to the case of just one modulus. The one-parameter family of metrics we get in this way can be written as:
\begin{equation}\label{EH}
ds^2_{EH}~=~\left( 1-\left(\frac{a}{r}\right)^4\right)^{-1}dr^2~+~r^2\left(1-\left(\frac{a}{r}\right)^4\right)\sigma_3^2~+~r^2(\sigma_1^2+\sigma_2^2),
\end{equation}
where $a\leq r\leq \infty$, $a^2$ is proportional to the size of the blown-up cycle, the $\sigma_i$ are defined in terms of the ${\mathbf S}^3$ Euler angles ($\theta,~\phi,~\beta$):
\begin{eqnarray}
2\sigma_1&=&-\sin\beta d\theta+\cos\beta\sin\theta d\phi,\\ \label{sigma}2\sigma_2&=&\cos\beta d\theta+\sin\beta\sin\theta d\phi,\\ 2\sigma_3&=&d\beta+\cos\theta d\phi
\end{eqnarray}
and they satisfy $d\sigma_a=\varepsilon_{abc}\sigma_b\wedge\sigma_c$. Note also that $4(\sigma_1^2+\sigma_2^2)$ is the standard unit radius ${\mathbf S}^2$ metric and $\sigma_1^2+\sigma_2^2+\sigma_3^2$ is the ${\mathbf S}^3$ one. As the angular variables have periodicity $0<\theta\leq\pi,\ 0<\phi,~\beta\leq 2\pi$, this space at infinity looks like $\mathbf{S}^3/\mathbb{Z}_2$.

The solution corresponding to fractional branes in the Eguchi-Hanson background has been found in \cite{Bertolini:2001ma} and it is very similar to the orbifold one. It can be written in terms of a metric
\begin{equation}\label{EHm}
ds^2=~H(r,\rho)^{-1/2}\eta_{\alpha\beta}dx^{\alpha}dx^{\beta}~+~H(r,\rho)^{1/2}\left(d\rho^2+\rho^2d\theta^2\right)~+~H(r,\rho)^{1/2}ds^2_{EH},\end{equation} a self dual R-R five-form\begin{equation}F_5~=~d\left(H(r,\rho)^{-1}~dx^0\wedge\ldots\wedge dx^3\right)~+~\star d\left(H(r,\rho)^{-1}~dx^0\wedge\ldots\wedge dx^3\right),\end{equation} and a complex three form (a linear combination of the NS-NS one and the R-R one) valued only in the transverse space:\begin{equation}\label{EH3} {\mathcal H}~\equiv~F_3^{NS}+iF_3^{R} ~=~d\gamma(\rho,\theta)\wedge\omega,\end{equation}
where $\gamma(\rho,\theta)$ is an analytic function on the transverse $\mathbb{R}^2$, $\omega$ is the harmonic anti-self-dual form on the Eguchi-Hanson space and $r$ is now the radial Eguchi-Hanson coordinate.

Even if the microscopic interpretation of this solution is quite obscure \cite{Bertolini:2001ma}, we assume that it corresponds to the standard geometrical interpretation of fractional $D3$-branes as $D5$-branes wrapped on the non-trivial cycle that shrinks to zero size in the orbifold limit \cite{Douglas:1996xg}. The form of the solution is indeed consistent with the interpretation of twisted fields as fields coming from wrapping of higher degree forms \cite{Douglas:1996xg}, as it was already confirmed in \cite{Merlatti:2000ne} by explicit computations of string scattering amplitudes. The main difference with respect to the orbifold solution is the explicit functional form of the warp factor $H(r,\rho)$, whose physical implication is that there is still a naked singularity screened by an enhan\c{c}on mechanism, but contrary to the orbifold case, in the internal directions (those along the ALE space) the singularity is cured by the blowing-up of the cycle and there is no enhan\c{c}on there \cite{Bertolini:2001ma}.

Unluckily, the Eguchi-Hanson background lacks of a conformal field theory description. This make it difficult to determine the correct form of the boundary actions one would need to get gauge theory information. It looks nevertheless quite reasonable to assume the proper boundary action is the same as the orbifold one with the inclusion of the new FI term corresponding to having blown up the cycle. This can be motivated by the knowledge of the explicit solution in the Eguchi-Hanson case, that makes it clear the close similarity to its orbifold limit. Besides, this is very analogous in spirit to the successful assumption made in \cite{Polchinski:1996ry}. The potential the brane fills will be thus the same we computed in the orbifold (we refer to the magnetic case (\ref{potm})) but now with an extra term corresponding to have $\xi\neq 0$ and proportional to the size of the blown-up cycle ($a^2$):
\begin{equation}
V(\Phi)~=~m^2 \tau_2+\frac{\xi^2}{\tau_2}\label{potem}
\end{equation}
We see again that the no-force condition is violated but now there is an interesting minimum where the probe-brane can sit, i.e. at \begin{equation}\rho_*~=~\rho_e e^{\frac{\pi}{N}\frac{\xi}{m}},\label{minimum*}\end{equation}where the enhan\c{c}on radius $\rho_e$ has been defined in (\ref{re}). As $\rho_*>\rho_e$, $\rho=\rho_*$ defines a regular space-time locus. From the field theory analysis we know this is a vacuum of the first type we discussed in section (\ref{vacua}) and as such it preserves just half of the supersymmetry. This means that the gauge theory living on the probe is ${\cal N}=1$ supersymmetric with a massless vector multiplet and a massive chiral one, with mass $M$ given by:\begin{equation}\label{mass}M=\frac{m^2}{2\xi}\langle\tau'\rangle.\end{equation} From the supergravity point of view we have again a solution with bulk ${\cal N}=2$ supersymmetry (as shown in \cite{Bertolini:2001ma}) but the embedding of the brane probe seems to break half of it. It would be interesting to see this purely in a stringy context by making a proper K-supersymmetry study of this geometry. However, our field theory analysis already shows that the probe-brane sees a geometry that is locally ${\cal N}=1$ supersymmetric. 

In the T-dual IIA set-up the chiral adjoint scalar gets a mass if the two NS5-branes are relatively rotated \cite{Barbon:1997zu}. As this type IIA geometry is eventually dual to the conifold one \cite{Uranga:1998vf}, we expect the conifold geometry to appear now.

\subsection{Resolved conifold}

We want now to analyze better the geometry seen by the probe brane. We start noting that, according to the interpretation of fractional $D3$-branes as $D5$-branes wrapped on the non trivial two-cycle of the Eguchi-Hanson space, the probe-brane naturally tends to go toward the point of the Eguchi-Hanson space where the cycle has minimal area, {\it i.e.} towards $r\to a$ in (\ref{EH}). Even if the point $r=a$ looks a singularity in the metric (\ref{EH}), it is well known that the limit $r\to a+\epsilon$ for small $\epsilon$ is well defined and it gives:
\begin{equation}
ds^2~\sim~\frac{1}{4}\left\{\frac{a}{\epsilon}\left[d\epsilon^2+4\epsilon^2(d\beta+\cos\theta d\phi)^2\right]+a^2(d\theta^2+\sin^2\theta d\phi^2)\right\},\end{equation}
which is topologically looking locally as ${\mathbb R}^2\times {\mathbf S}^2$ (globally there is a further fibered structure due to the $d\beta d\phi$ cross term). To see this more clearly, we make the further change of coordinate $\epsilon=v^2/a$ and the metric becomes:
\begin{equation}\label{EHr}
ds^2~=~dv^2~+~v^2(d\beta+\cos\theta d\phi)^2~+~\frac{a^2}{4} (d\theta^2+\sin^2\theta d\phi^2).
\end{equation}
This makes it clear as, having $\beta$ periodicity $2\pi$, the putative singularity at $r=a$ is simply an artifact of the coordinates. 

In the ${\mathbb R}^2$ directions (parameterized by $\rho$ and $\theta$ in (\ref{EHm})) the probe-brane sits instead at the circle defined by $\rho=\rho_*$ and the value of $\rho_*$ (given in (\ref{minimum*})) varies with the parameters of the solution. 

Finally, the geometry seen by the probe-brane is determined by the ${\mathbb R}^2\times EH$ metric, in the $r\to a$ limit. This space looks locally as ${\mathbb R}^4\times{\mathbf S}^2$:
\begin{equation}
ds^2~=~dx^2~+~dy^2~+~dv^2~+~v^2(d\beta+\cos\theta d\phi)^2~+~\frac{a^2}{4} (d\theta^2+\sin^2\theta d\phi^2), \end{equation}
where now we have chosen to parameterize ${\mathbb R}^2$ with Cartesian coordinate $x$ and $y$. Making now the coordinate redefinition: 
\begin{eqnarray}
x~=~u \cos \frac{\tilde{\theta}}{2}\cos\left(\frac{\alpha+\tilde{\phi}}{2}\right),\\
y~=~u \cos \frac{\tilde{\theta}}{2}\sin\left(\frac{\alpha+\tilde{\phi}}{2}\right),\\
v~=~u\sin\frac{\tilde{\theta}}{2},\\
\beta~=~\frac{\alpha-\tilde{\phi}}{2},
\end{eqnarray}
where $0<\alpha\leq 4\pi,\ 0<\tilde{\phi}\leq 2\pi,\ 0<\tilde{\theta}\leq\pi$, we get
\begin{eqnarray}\nonumber
ds^2~=~du^2~+~\frac{u^2}{4}\big[(d\alpha+\cos\tilde{\theta}d\tilde{\phi}+\cos\theta d\phi)^2+d\tilde{\theta}^2+\sin^2\tilde{\theta}d\tilde{\phi}^2\\-2\cos\theta d\phi(d\tilde{\phi}+\cos\tilde{\theta}d\alpha)\big]~+~\frac{a^2}{4} (d\theta^2+\sin^2\theta d\phi^2),\label{EHnh}
\end{eqnarray}
where terms involving higher power of $u$ with two differentials in the $\theta,~\phi$ directions have been neglected. As shown in the Appendix, this metric is locally the same as the ``near horizon'' of the resolved conifold one (see eq. (\ref{resolvedr}) in the appendix). We note that globally there is a small difference between the two, being the fibration of the blown-up ${\mathbf S}^2$ not identical (due to the cross term $\cos\theta d\phi(d\tilde{\phi}+\cos\tilde{\theta}d\alpha)$ in (\ref{EHnh})). 

It is perhaps not surprising we find the resolved conifold geometry appearing here: it is indeed known the resolved conifold geometry is the proper one to engineer ${\cal N}=1$ pure supersymmetric Yang-Mills theory in the type IIB set-up \cite{Vafa:2000wi}. In this section we have just followed this engineering entirely in terms of supergravity solutions and we clearly see its relation with the parent ${\cal N}=2$ orbifold theory. A non-trivial information we immediately have (with the help of the field theory analysis and particularly thanks to the formula (\ref{mass})) is that the size of the blown-up two-cycle is directly related to the inverse mass of the chiral multiplet. 

The assumption about the microscopic interpretation of the probe as wrapped $D5$-branes cannot be proven here, but, as we will see, it makes all the picture consistent. Under this assumption, we have learned that, if a $D5$-brane is wrapped on the non-trivial two-cycle of a resolved conifold, close to the apex of such cone the gauge theory living on its world-volume is ${\cal N}=1$ supersymmetric. The embedding of $N$ branes in such background will then give rise to ${\cal N}=1$ supersymmetric $U(N)$ Yang-Mills theory. The fact we reach such conclusion via a probe analysis means that we are studying open strings (and then branes) in the given background. This matches perfectly with the picture of geometric transition we discussed in the introduction: being talking of open strings and branes, we are before the transition and accordingly the background is the resolved conifold! 

From \cite{Vafa:2000wi,Klebanov:2000hb}, we know how the story goes on, but we briefly review it here following the perspective we have just developed. To get pure ${\cal N}=1$ super-Yang-Mills, we take the zero size limit of the blown-up cycle. In this way the chiral adjoint field acquires infinite mass (see eq. (\ref{mass})) and it is decoupled. The relevant geometry is now the conifold one. This geometry is still singular \cite{Klebanov:2000nc} and according to equation (\ref{minimum*}) the brane will tend to go to the enhan\c{c}on locus. Consistently, it has been shown that at least from a supergravity point of view the enhan\c{c}on is a consistent mechanism also in the conifold case \cite{Merlatti:2001gd}. In principle there are two ways now to de-singularize the geometry (or we could also say to determine the geometry inside the enhan\c{c}on shell): the $\mathbf{S}^2$ or the $\mathbf{S}^3$ can be blown-up. As we don't want to go back to the original case, the only remaining possibility is to look for a consistent deformed conifold solution. From \cite{Klebanov:2000hb} we know this is indeed the right thing to do: the deformed conifold is a non-singular solution and its geometry is the proper one to describe the infrared ${\cal N}=1$ Yang-Mills dynamics.

\section{Conclusions}

In this paper we have described how partial supersymmetry breaking can be obtained in the context of the gauge/string correspondence. This has been reached studying the low energy dynamics of open strings in given backgrounds. The tool we used to perform such study is the probe-analysis. We have seen in this way that a potential, due to the internal flux, is generated in the ${\cal N}=2$ orbifold. This potential drives the probe-brane to the enhan\c{c}on locus. Instead, when the orbifold singularity is softened to the smooth Eguchi-Hanson space, the resulting potential drives the probe-brane to regular space-time points, ({\it i.e.} outside the enhan\c{c}on shell). We have found that at those points (corresponding to minima of the potential) partial supersymmetry breaking does occur. We have further analyzed the geometry seen by the probe-brane close to those minima and we have found it is the resolved conifold one, in the vicinity of the blown-up two-cycle. We have finally showed that shrinking the two-cycle to zero size corresponds, on the field theory side, to decouple the adjoint chiral scalar and thus to get pure ${\cal N}=1$ super-Yang-Mills. 

An issue deserving more investigation is that our findings seem to suggest that the T-dual type IIA configuration of rotated NS5-branes corresponds, for generic angle of rotation, to the resolved conifold geometry. The singular conifold would be the relevant one just for the case of perpendicular NS-branes.

\vskip 1cm \centerline{\bf Acknowledgments} \noindent I thank M. Bill\`o, P. Di Vecchia, M. Frau, G. Vallone and especially A. Lerda for many useful discussions and exchange of ideas. This work is partly supported by the EU contracts MRTN-CT-2004-503369 and MRTN-CT-2004-512194, DAAD, and the DFG grant SA 1336/1-1.

\appendix
\section{Resolved conifold metric}

We review here briefly the resolved conifold case \cite{PandoZayas:2000sq}\footnote{see also \cite{Becker:2004qh} for a way to embed the non-supersymmetric solution of \cite{PandoZayas:2000sq} in a supersymmetric set-up}. Its metric is: 
\begin{eqnarray}
ds^2~=~K^{-1}(u)du^2~+~\frac{1}{9}K(u)u^2(d\psi+\cos\theta_1d\phi_1+\cos\theta_2d\phi_2)^2~+\\ \frac{1}{6}u^2(d\theta_1^2+\sin^2\theta_1d\phi_1^2)\nonumber+\frac{1}{6}(u^2+b^2)(d\theta_2^2+\sin^2\theta_2d\phi_2^2),\end{eqnarray}
where \begin{equation} k(u)~=~\frac{u^2+9b^2}{u^2+6b^2}.\end{equation}
For small $u$ we get now: \begin{eqnarray}\label{resolvedr} ds^2&\sim&\frac{2}{3}\Bigg[du^2+\frac{1}{4}u^2(d\psi+\cos\theta_1d\phi_1+\cos\theta_2d\phi_2)^2\\&&+\frac{1}{4}u^2(d\theta_1^2+\sin^2\theta_1d\phi_1^2)~+~ \frac{b^2}{4}\left(d\theta_2^2+\sin^2\theta_2d\phi_2^2\right)\Bigg]\nonumber\end{eqnarray}

We note finally that this is locally the same metric as $\mathbb{R}^2\times EH$ near the two cycle (\ref{EHnh}). Globally there is a small difference between the two, as the fibration of the two ${\mathbf S}^2$ are not identical (see the last ``cross'' terms in (\ref{EHnh})).

\end{document}